\documentclass[a4paper,twocolumn]{article}
\usepackage{graphicx}
\usepackage{cite}
\usepackage{microtype}
\usepackage{float}
\usepackage{titlesec}
\usepackage{paralist}

\usepackage[english]{babel}
\renewcommand\thesection{\Roman{section}}
\titleformat{\section}[block]{\large\scshape\centering}{\thesection.}{1em}{}

\date{}

\begin{document}

\title{\fontsize{14pt}{12pt}\selectfont\sc Amplification and passing through the barrier of the exciton condensed phase pulse in double quantum wells}

\author{\normalsize O.\,I.~Dmytruk$^1$ and V.\,I.~Sugakov$^2$\\
\\\normalsize
\texttt{$^2$sugakov@kinr.kiev.ua} \vspace*{-8mm}}

\twocolumn[
\begin{@twocolumnfalse}
\maketitle

The peculiarities and the possibility of a control of exciton condensed pulse movement in semiconductor double quantum well under the slot in the metal electrode are studied. The condensed phase is considered phenomenologically with the free energy in Landau-Ginzburg form taking into account the finite value of the exciton lifetime. It was shown that the exciton condensed phase pulse in the
presence of an external linear potential moves along the slot direction 
with a constant value of a maximum density during exciton lifetime, while the exciton gas phase pulse is blurred. The penetration of the exciton condensed phase pulse through the barrier and its stopping by the barrier are studied. Also, it was shown that the exciton pulse in the condensed
phase can be amplified and recovered after damping by imposing an additional laser pulse. Solutions for the system of excitons in double quantum well under the slot in the electrode under steady-state irradiation in the form of bright and dark autosolitons were found.

~


~

\end{@twocolumnfalse}
]

\section{Introduction}
In recent years, much attention has been paid to both experimental and
theoretical study of indirect excitons in semiconductor double quantum wells at
low temperatures. An indirect exciton is a bound pair of an electron and a hole
which are separated by an electric field to different quantum wells~\cite{fukuzawa1990phase}.
Consequently, the recombination of the electron and the hole is inhibited, that
causes long lifetime of indirect excitons, which is several orders of
magnitude higher than the direct exciton lifetime. The study of indirect
excitons is promising both in terms of fundamental science, because a high
density of excitons can be created and many
exciton effects can be studied. Also the system of indirect excitons can be promising for application purposes, since they can
travel over large distances, carry energy and information and be used in
semiconductor devices based on double quantum wells. Therefore, many
experimental and theoretical works were
devoted to the study of the
indirect excitons' properties. Under these investigations, the experimenters on the base of $AlGaAS$ quantum wells~\cite{larionov2000interwell, dremin2004bose, timofeev2005collective, butov2002macroscopically, snoke2002long, gorbunov2006large, gorbunov2006bose, timofeev2011bose} revealed new nontrivial effects such as
formation of spatially inhomogeneous structures (sometimes periodical) in the
distribution of exciton density, and the fact that symmetry of the formed
structures is not connected to the symmetry of the external field. In
theoretical works~\cite{levitov2005pattern, paraskevov2007microscopic, saptsov2008instability, liu2006pattern, mukhomorov2010possibility, sugakov2005islands, chernyuk2006ordered,  sugakov2007formation, sugakovc2007formation, sugakov2009ordered,   wilkes2012drift, andreev2012thermodynamic, babichenko2013many} various models of pattern formation were proposed. However,
as the interpretation of the results in different works is versatile, so research in this
direction is relevant. The possibility of using the system of indirect excitons
in the optoelectronics is studied in several works~\cite{high2007exciton, high2008control, winbow2011electrostatic, remeika2009localization, kuznetsova2010all}. Namely, the
experimental possibility of building an exciton optoelectronic transistor~\cite{high2007exciton},
an excitonic integrated circuit~\cite{high2008control} and an excitonic conveyor~\cite{winbow2011electrostatic} were
demonstrated recently. Therefore, the theoretical study of the movement of the
interacting system of indirect excitons in external fields is important. 

In this work we study the formation, amplification and passing through the barrier of the
exciton condensed phase pulse in the semiconductor double quantum well in the system, in which one of the electrodes has a slot, which creates an additional field to uniform field for excitons. The
formation of the exciton condensed phase in the double quantum well under the slot excited by a steady-state uniform irradiation was studied in~\cite{sugakov2009ordered}. The approach used to describe the
distribution of the exciton density in double quantum wells is based on the
assumption of the exciton condensed phase existence due to the attraction
between indirect excitons at some distances and the finite value of the exciton lifetime. The existence of the attractive
interaction between excitons is confirmed by the calculations of many- exciton
system~\cite{lozovik1996phase} and by the possibility of biexciton formation~\cite{tan2005exciton, schindler2008analysis, meyertholen2008biexcitons}.  Also, the
finite value of the exciton lifetime plays an important role in the formation
of the structures in the exciton density distribution. The suggested model explained a
series of experimental results (see references in~\cite{sugakov2009ordered}).  Particularly, this
model has allowed to describe spatial distribution of the excitons on the ring
outside the laser spot, observed in~\cite{butov2002macroscopically}, spatial structure of the luminescence
from the double quantum well under the window in the metal electrode, observed
in~\cite{gorbunov2006bose, gorbunov2006large}.

Unlike the work~\cite{sugakov2009ordered}, where the structure of the exciton density distribution
in the quantum well under the slot in the case of uniform in space and time
independent pumping was studied, this paper investigates the excitation of the
exciton condensed phase by pulse irradiation. This excitation creates the
exciton condensed phase pulse, which movement and control are analyzed
in this paper. The possibility of the exciton condensed phase pulse
amplification by additional laser pulse in the one-dimensional system was
investigated in~\cite{dmytruk2012movement}. In comparison to~\cite{dmytruk2012movement}, we consider the real physical system (slot in the electrode)
with two-dimensional distribution of the exciton density, so the two-dimensional problem for the exciton density is studied. Movement
of the pulse takes place in the quantum well under the slot and along it. And along with amplification of the exciton condensed phase pulse, we consider passing of the exciton condensed phase through a barrier. Presence of the barrier allows to control exciton pulse, to stop it in particular. A consideration of the processes with exciton system in the condensed phase means that we are studying the collective effects of exciton system movement and passing through barrier.

\section{Model of the system}
\begin{figure}[h!]
	\centering
	\includegraphics[width=\linewidth]{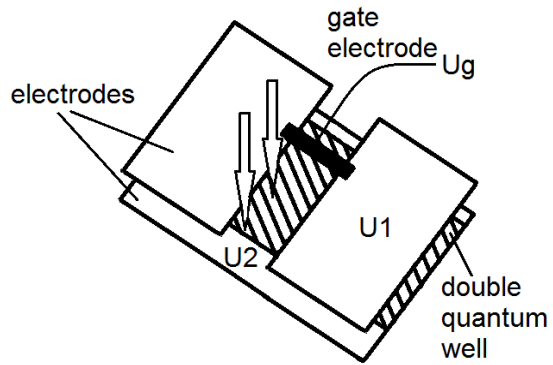}
	\caption{Pattern of the system. Semiconductor structure sandwiched between two metal
electrodes. There is a slot in the upper electrode. $U1$ is the potential on the
upper electrode, $U2$ is the potential on the lower electrode and $Ug$ is the gate
potential of the strip. Perpendicular metal strip that creates a barrier for
excitons. Arrows show laser irradiation of the double quantum well.}
	\label{fig:figure1}
\end{figure}

Let us consider the semiconductor structure with the double quantum well sandwiched
between two metal electrodes. There is the slot in the upper electrode (figure~\ref{fig:figure1}). The
width of the slot is $2b$. Let us direct the OX axis along the slot, the axis OY in the transverse
direction and the OZ axis is directed along the normal to the electrode.  Let the plane of the quantum well has a coordinate
$z$ $(z<0)$. The presence of the slot forms an inhomogeneous electric field
that is additional to the homogeneous field of solid electrodes and an
additional potential energy for excitons in quantum well.

In~\cite{sugakov2009ordered} the existence of a periodic distribution of the exciton density under a
homogeneous stationary irradiation was shown. If the width of the slot is large,
then a string of the islands of the exciton condensed phase is divided into two
lines with the islands of the exciton condensed phase located along the slot
edges and shifted relative to each other in rows along the axis of the
slot in the half-period. Periodic structure exists in a certain range of
pumping values. Also in~\cite{sugakov2009ordered} the possibility of periodic structure movement in
the presence of an external potential (dependent on the coordinates) along the
slot was shown. The movement is analogous to exciton Gunn effect in semiconductors.
In the considered system, in comparison to the work~\cite{sugakov2009ordered}, there are additional elements, that are pointed in the figure~\ref{fig:figure1} and they will be explained later. These elements allow to control exciton pulses.

The potential energy $V_{tot}$ for exciton in quantum well, created by electrodes, may be presented as $V_{tot}=-p_zE(y,z)$, where $p_z$ is the exciton dipole moment, $E(y,z)$ is the electric field strength in the $y,z$  point of quantum well. This electric field is uniform in the region far from the slot $E(y,z)=E_0$. Additional potential for excitons, created by slot, equals $V(x,y)=V_{tot}-V_0$, where $V_0=-p_zE_0$ is the uniform shift of the exciton band under the electric field in the region distant from the slot. Using the solution for the electric field of a grounded conducting plane with a
slot in an external electric field problem~\cite{landau1998electrodynamics}, and making the
assumption that the distance between the electrodes is much greater than the
width of the slot, for additional potential energy for exciton formed by the
slot, we obtain the following equation:
\begin{equation}
\begin{array}{ll}
V(y,z) =& V_0 - V_0 \bigg[ (\frac{1}{2} + \frac{1}{2}\sqrt{1 + b^2/\xi(y, z)}) - \\
        & -\displaystyle\frac{b^2z^2}{2\xi(y, z)^2\sqrt{1 + b^2/\xi(y, z)}} \cdot \\
        & \cdot \Big(1 + \displaystyle\frac{y^2 + z^2 + b^2}{\sqrt{(y^2 + z^2 - b^2)^2 +
          4z^2b^2}} \Big) \bigg],
\end{array}
\label{equ:adpotential}
\end{equation}
\noindent where 
\begin{equation}
\begin{array}{ll}
\xi(y, z) =& \frac{1}{2} \Big[ y^2 + z^2 - b^2 + \cr
         &+\sqrt{(y^2 + z^2 - b^2)^2 + 4b^2z^2} \Big].
\end{array}
\end{equation}

\begin{figure}[h!]
  \centering
 \includegraphics[width=\linewidth]{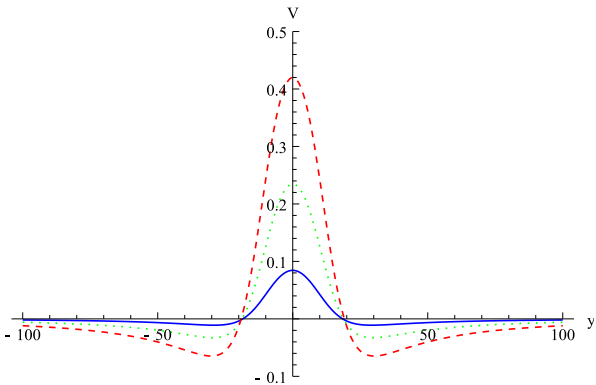}
  \caption{The dependence of the potential created by the slot on the distance
from the centre of the slot in the plane of the quantum well for dimensional units that will be introduced in the next section $V_0=5$, $z=-15$ at different width of the slot: 1) $b=10$ (dashed curve), 2) $b=7$ (dotted curve), 3) $b=4$ (solid curve).}
\label{fig:figure2}
\end{figure}

The potential in the quantum well under the slot has the maximum and is
positive at the centre of the slot (figure~\ref{fig:figure2}), because the electric field in this regions is less than the field far from the slot. But in a certain vicinity of the borders of the slot the potential
has a small minimum with a negative value. This appears due
to the rearrangement of charges on the conductive electrode in
the vicinity to edges of the slot. The depth of the minimum increases as
the quantum well approaches to the electrode with a slot.

\section{Method of solution}
The studied system will be described more detailed in the next section, but in this section we will concentrate our attention on the main equations for finding the density dependence of the exciton on time and space.
For investigation of the exciton distribution let us consider
the conservation law for the exciton density as follows:
\begin{equation}
 \displaystyle\frac{\partial n}{\partial t}  = - \mathrm{div} \vec{j} +
 G(\vec{r},t) -\displaystyle\frac{n}{\tau},
 \label{equ:pumping_derivative}
\end{equation}
\noindent where $n$ is the exciton density, $G(\vec{r},t)$ is the pumping (the number of
excitons created per time unit per space unit of a quantum well), $\tau$ is the
exciton lifetime, $\vec{j} = -M \vec{\nabla} \mu$ is the exciton current
density, $M$ is the exciton mobility. 
For the exciton mobility $M$ we use Einstein formula 
$M = \displaystyle\frac{n D}{k T}$, where $D$ is a diffusion coefficient, $k$ is
Boltzmann constant, $T$ is the temperature.
Described system is non-equilibrium because the pumping is present, and during
phase transitions the system is also inhomogeneous. But the time for
establishment of equilibrium in a local area is much less than in the whole
system, because the equilibrium in the whole system is formed by slow diffusive
processes. That is why the system can be described by free energy that depends
on exciton density, which in its turn depends on the space coordinates.

Having expressed the chemical potential as $\mu = \displaystyle\frac{\delta F}{\delta n}$,
we chose the free energy in the Landau model:
\begin{equation}
  F[n] = \int \mathrm{d} \vec{r} \left[
    \displaystyle\frac{K}{2} {\left( \vec{\nabla} n \right)}^2 + f(n) + n V
  \right].
  \label{equ:energy}
\end{equation}
The term $\displaystyle\frac{K}{2} {\left( \vec{\nabla} n \right)}^2$ characterizes the
energy of inhomogeneity, $f(n)$ is a free energy density, $nV$ accounts for the
potential energy.  We approximate the density of the free energy in the form:
\begin{equation} 
f(n) = f_0 + kTn(\ln\displaystyle\frac{n}{n_0} - 1) + \displaystyle\frac{a}{2} n^2 + \displaystyle\frac{b}{3} n^3 +
\displaystyle\frac{c}{4} n^4;
\label{equ:dens}
\end{equation}
\noindent where the terms of the power series for the exciton density are important for large
$n$, they are the most important for describing the system, while the term $k T n
(\ln\displaystyle\frac{n}{n_0} - 1)$ may be essential only if one wants to describe a
system in which the density of excitons is small and the interaction between
them is not important. 

Let introduce dimensionless variables for length,
density, energy and time:
\begin{equation}
  l_u = \sqrt{\displaystyle\frac{K}{a}},
  n_u = \sqrt{\displaystyle\frac{a}{c}},
  V_u = an_u,
  t_u = \displaystyle\frac{d_1l_u^2}{D}.
\end{equation}

Introducing the dimensionless variables and substituting equation~(\ref{equ:energy})
with equation~(\ref{equ:pumping_derivative}), we obtain a nonlinear equation that
determines the exciton density:
\begin{equation}
\begin{array}{ll}
\displaystyle\frac{\partial n}{\partial t} =& d_1\Delta n + \vec{\nabla} \Big[ n \vec{\nabla} (-\Delta n + n + \\
  & + b_1 n^2 + n^3 + V) \Big] + G - \displaystyle\frac{n}{\tau} -g n^2, \\
\end{array}
\label{equ:density_derivative_nodim}
\end{equation}
\noindent where $b_1 = \displaystyle\frac{b}{\sqrt{a c}} <0 $,
$d_1 = \displaystyle\frac{k T}{V_u} $ and
$V$ is an external potential, dependent on the coordinates.

Exciton-exciton annihilation (Auger process) was taken into account by introducing in equation~\ref{equ:density_derivative_nodim} the term
$-g n^2$.

Equation~\ref{equ:density_derivative_nodim} is non-linear phenomenological equation that describes the
distribution of exciton density with regard to the pumping $G(\vec{r}, t)$ and
finite lifetime $\tau$.

\section{Creation and movement of the exciton condensed phase pulses}
Exciton pulse can be formed under the irradiation of the system by a laser
pulse. The laser pulse creates the following pumping for excitons:
\begin{figure*}[ht!]
\begin{equation}
G(x,y,t)=\left\{ \begin{array}{ll} P\exp\Big(-\frac{(x-x_0)^2+(y-y_0)^2}{\beta^2}\Big) & \mathrm{if~} t_1 < t < t_1 + \Delta t, y - b < y < y + b ; \\
                  0 & \mathrm{~in~all~other~cases}.\\
         \end{array} \right.
\label{equ:lpumping}
\end{equation}
\end{figure*}
\noindent (in the formula for the laser pulse, the electrodes non-transparency and an
additional constraint on the pulse of the y coordinate are taken into account).

$P$ is proportional to number of excitons created by the laser pumping per time unit per space unit of a quantum well. The laser pulse causes the temperature increasing. Considering this problem we assume that time of establishment equilibrium state of temperature is much less than the exciton lifetime $\tau$.

\begin{figure}[h!]
\includegraphics[width=\linewidth]{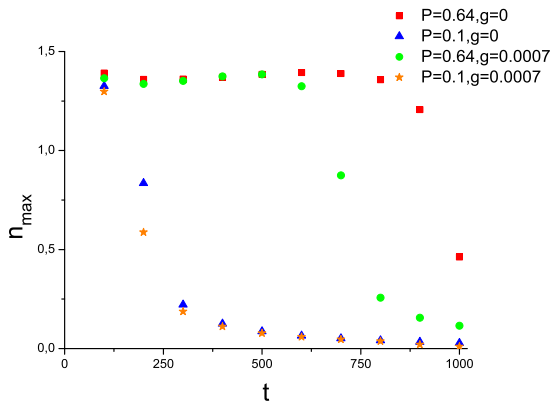}
  \centering
  \caption{Dependence of the exciton maximum on time for different values of
the parameters. Squares correspond to the exciton pulse in the condensed phase
$(P=0.64,g=0)$, triangles correspond to the exciton pulse in the gas phase
$(P=0.1,g=0)$, circles correspond to the exciton pulse in the condensed phase
taking into account Auger recombination of the excitons $(P=0.64,g=7\cdot 10^{-4})$,
stars correspond to the exciton pulse in the gas phase taking into account
Auger recombination of the excitons $(P=0.1,g=7\cdot 10^{-4})$. $b=4$,  $\beta=10$, $\tau=1000$, $b_1=-2.23$, $d_1=0.2$}
\label{fig:figure3}
\end{figure}

First, let us consider changing of the exciton density in the pulse in the case of the uniform potential along the slot. If the value of the amplitude of the laser pumping ($P$) is
less than the threshold value, the exciton pulse is formed in the gas phase. In
this case, the maximum of the exciton density in the pulse decreases rapidly with
time after the end of the laser pulse acting on the system (figure~\ref{fig:figure3}), the width of the
exciton pulse increases. If the value of the amplitude of the laser pulse exceeds the
threshold value, the pulse is formed from excitons in the condensed phase. Maximum value of the exciton density in the pulse formed from
the condensed phase remains constant approximately during the exciton lifetime,
width of the pulse decreases with time. 

In the case of the slot width $b$ less than some value, the pulse is created in
the centre of the slot (equation~\ref{equ:lpumping}), where the pumping has maximum. For the large value of the width, the exciton pulse will be formed at the edge of the slot.

Exciton-exciton annihilation decrease the way that can be passed by the exciton pulse because the attenuation. The annihilation affects more effectively the duration of the
pulse maximum value conservation in the case of the condensed phase (figure~\ref{fig:figure3})). This effect
 is obviously caused by the fact that the exciton density in the condensed
phase is greater. During calculations of annihilation effect we have chosen a large annihilation rate such
that the lifetime of a relatively annihilation and radiation are the same.

\begin{figure*}[ht!]
  \vspace*{-3cm}
  \centering
  \begin{minipage}[b][7cm][b]{\linewidth}
    \begin{minipage}[t]{0.5\linewidth}
      \centering
      \includegraphics[width=\linewidth]{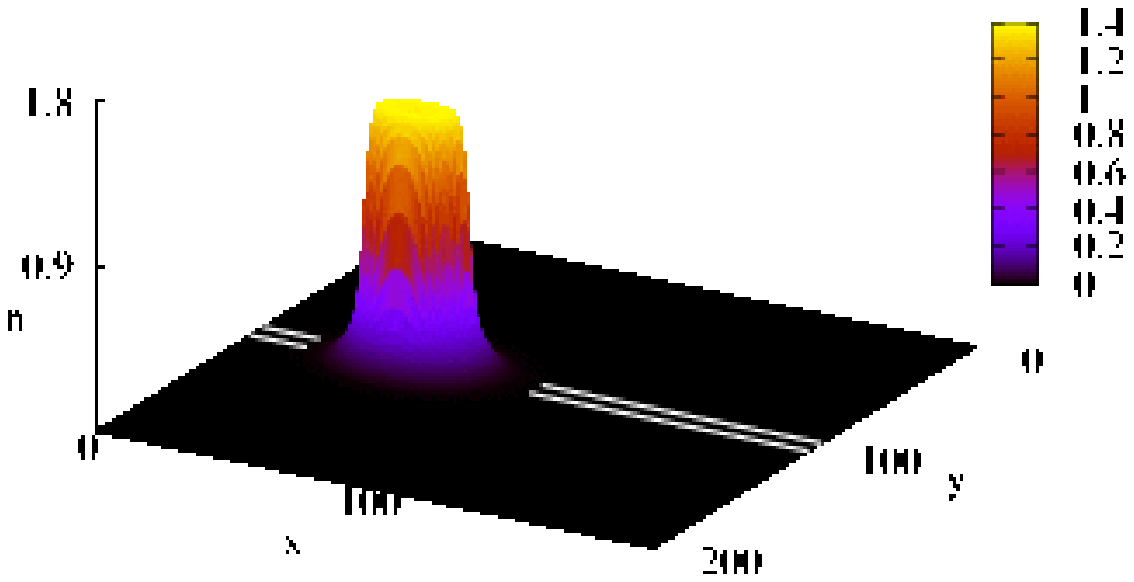}
      \par {\footnotesize {\bf a)} }
    \end{minipage}
    \hfill
    \begin{minipage}[t]{0.5\linewidth}
      \centering
      \includegraphics[width=\linewidth]{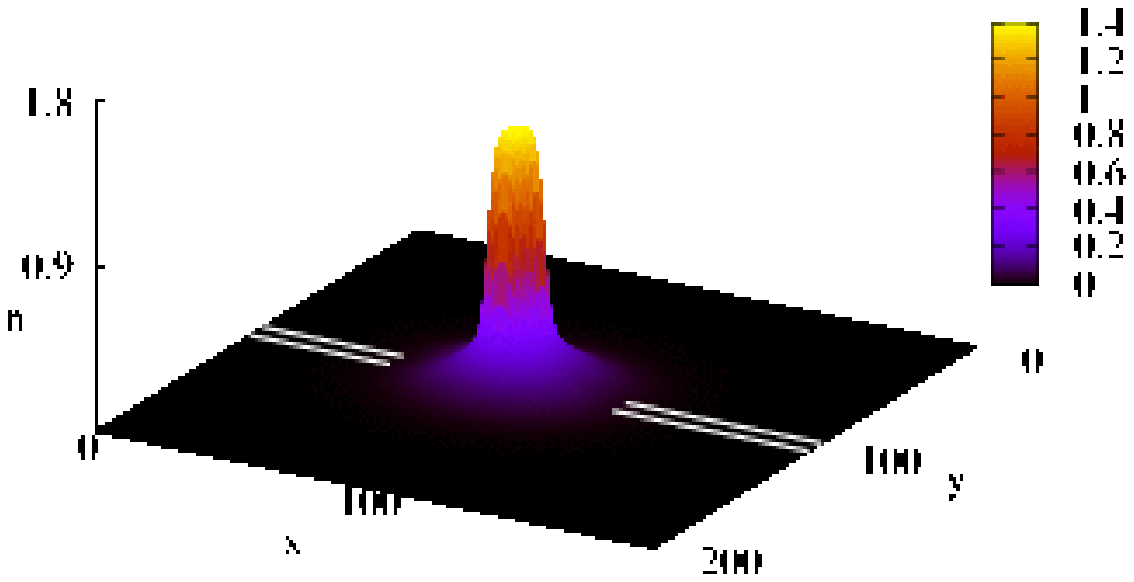}
      \par {\footnotesize {\bf b)} }
    \end{minipage}
  \end{minipage}
  \begin{minipage}[t][7cm][t]{\linewidth}
    \begin{minipage}[t]{0.5\linewidth}
      \centering
      \includegraphics[width=\linewidth]{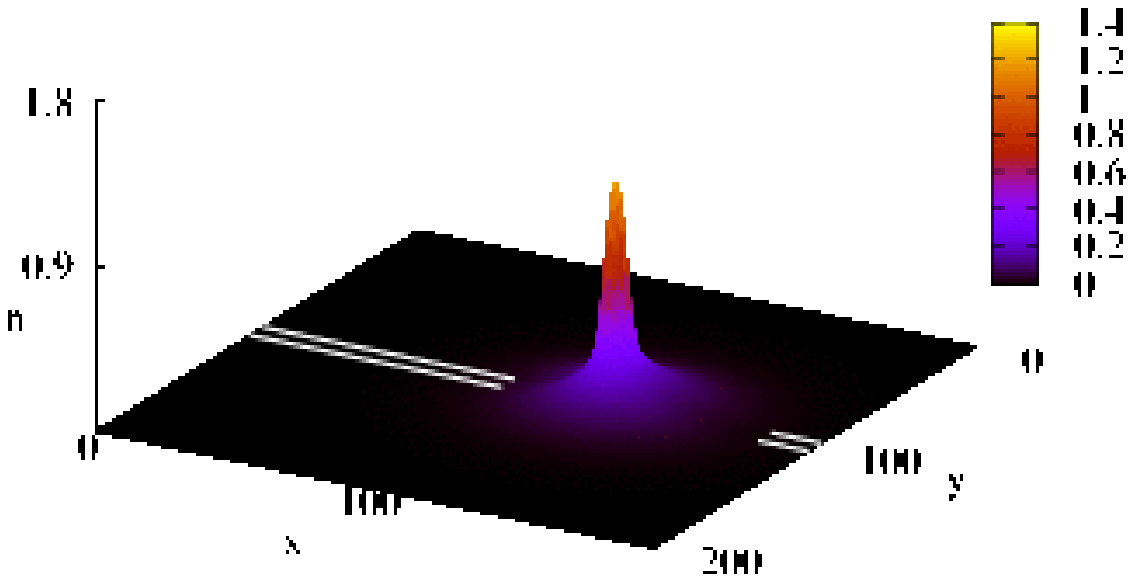}
      \par {\footnotesize {\bf c)} }
    \end{minipage}
    \hfill
    \begin{minipage}[t]{0.5\linewidth}
      \centering
      \includegraphics[width=\linewidth]{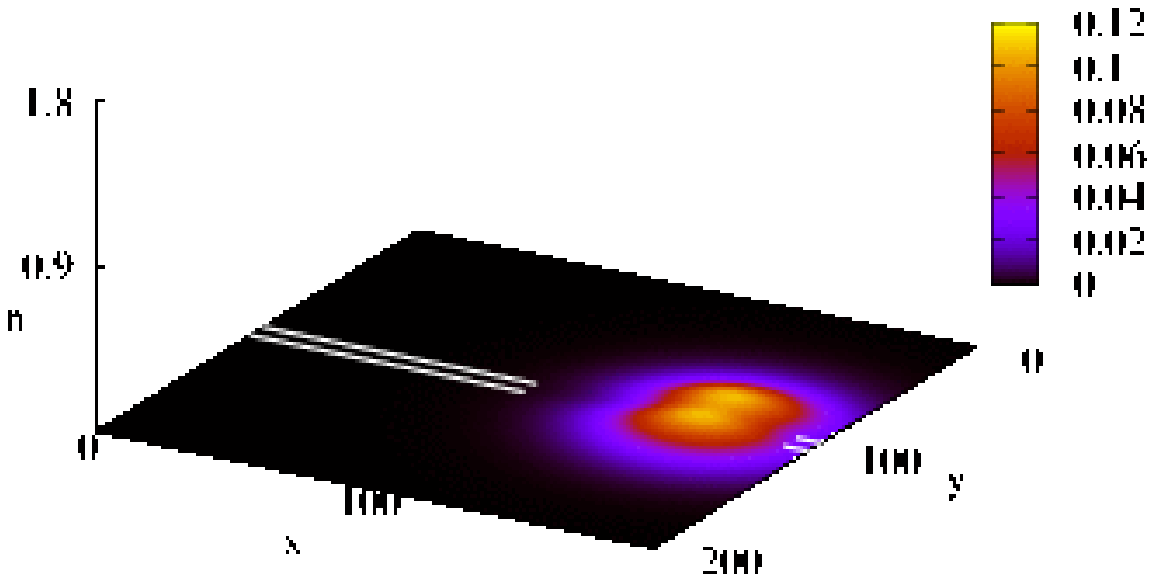}
      \par {\footnotesize {\bf d)} }
    \end{minipage}
  \end{minipage}
  \caption{Spatial distribution of the exciton density at $b=4$, $P=0.64$, $\beta=10$, $\tau=1000$, $b_1=-2.23$, $d_1=0.2$, $\delta=0.09$ at different moments of time after switching off the pumping:a) $t=100$, b) $t=500$, c) $t=900$, d)$t=1300$. The white lines show the edges of the slot.}
  \label{fig:figure4}
\end{figure*}

\begin{figure}[h!]
\includegraphics[width=\linewidth]{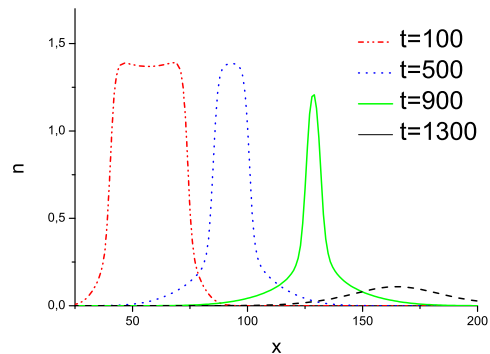}
  \centering
  \caption{Dependence of the exciton density on the $x$ coordinate at different moments of time without the barrier (parameters are the same as in figure~\ref{fig:figure4}).}
\label{fig:figure5}
\end{figure}

Let us consider the problem of the exciton pulse movement. To do this, let us
assume that beside the potential of the slot, there is an additional  potential
that depends linearly on the coordinates along the slot in the equation for the exciton
density.

\begin{equation}
V_l(x)=-\delta x,
\label{equ:linear}
\end{equation}

\noindent where $\delta$ is constant value.
This potential can be created, for example, by using an electric field, as it was
done by the authors~\cite{leonard2012transport}, or by an inhomogeneous stress of the crystal. The exciton pulse is presented in the figure~\ref{fig:figure4} at different moments of time. And the projection of density distribution on XZ plane is presented in the figure~\ref{fig:figure5}. One can see that
the exciton condensed phase pulse moves along the slot. The velocity of the exciton
pulse movement depends on the potential ($v \approx \displaystyle\frac{\delta D}{k T}$). It is seen from these figures, that the maximum value of the exciton density in the pulse does not change almost during its lifetime, if the exciton pulse is in the condensed phase. But the width of the pulse decreases with time. In the case of the exciton pulse in the gas phase, the exciton pulse blurs. 

In the case of the wide slot two exciton pulses are formed and they also
move along the slot in parallel to each other.

So one can use the exciton pulse
for data transmission in a semiconductor system. If the detector at the output
reacts to the amplitude of the exciton density, one can achieve greater
distances at which the pulse of excitons is in the condensed phase state.

\section{Amplification of the exciton condensed phase pulse by the laser pulse}
\begin{figure}[h!]
  \centering
\includegraphics[width=\linewidth]{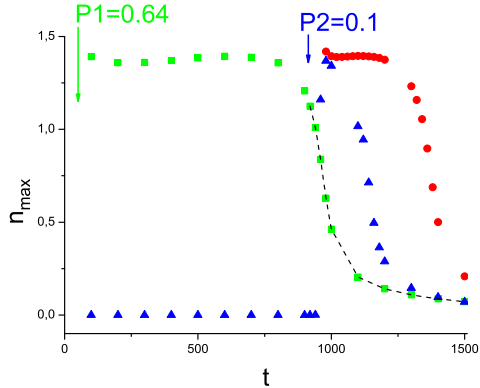}
  \caption{Dependence of the exciton maximum magnitude ($n_{max}$) on time ($t$). Squares correspond to the first pulse exciton pulse ($P=0.64$), triangles correspond to the second pulse ($P=0.1$),
circles correspond to the total pulse and dashed line correspond to the decreasing of
the maximum exciton density without additional laser pulse.}
\label{fig:figure6}
\end{figure}

Exciton condensed phase pulse can be amplified by imposing an additional laser
pulse on the system. Let us consider the exciton pulse in the condensed phase. As was shown already, the exciton density is constant during the exciton lifetime and after some time the amplitude of the pulse begins to drop and the system of  excitons in the pulse transform to excitons in the gas phase. To recover the exciton condensed phase pulse, let us apply at some fixed moment of time additional laser pulse on the system, amplitude of which can be lower the threshold value for creation of the exciton condensed phase pulse. But imposing the laser pulse on the exciton pulse, when the position by the imposing time matches the position
of the first pulse maximum, leads to creation of the exciton pulse in the condensed phase. Imposing of the two laser pulses arise the exciton pulse in the condensed phase, which exists for longer time and can travel over larger distance (dependence of the maximum value of the exciton density in the first, second and total pulses on time are presented in figure~\ref{fig:figure6}). If a detector reacts to the amplitude of the exciton pulse, then the signal being transmitted by the system in the condensed
phase can be received on a greater distance.

\section{Control of the exciton condensed phase pulse using the barrier}

\begin{figure}[h!]
\includegraphics[width=\linewidth]{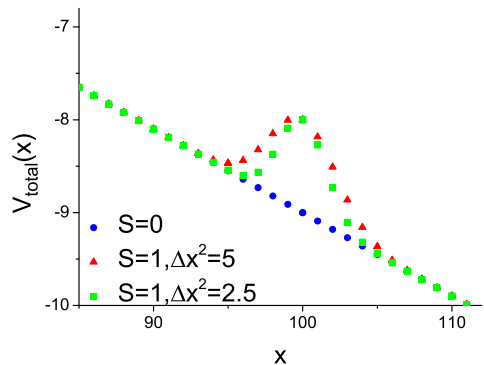}
  \centering
  \caption{Total additional potential for excitons created by linear potential and barrier ($V_l(x)+V_b(x)$), $\delta=0.09$, $S=1$.}
\label{fig:figure7}
\end{figure}

The exciton condensed phase pulse can be controlled, for example, using a barrier. The barrier can be created by placing a thin metal
strip perpendicularly to the slot (see figure~\ref{fig:figure1}). The barrier for excitons of
different magnitude can be formed by applying a potential to the strip. Let us
approximate the additional potential for excitons created by the strip as follows:

\begin{equation}
V_b(x)=S\exp(-(x-x_0)^2/2(\Delta x)^2),
\label{equ:barrier}
\end{equation}

\noindent where $S$ is the magnitude of the barrier and $\Delta x$ is the width of the barrier.  Cumulative
potential for excitons created by the linear potential (equation~\ref{equ:linear}) and the barrier (equation~\ref{equ:barrier})
is presented in figure~\ref{fig:figure7},  the parameters of these potentials are $\delta=0.09$, $S=1$.  One can observe that in some cases ($S=1$) an obstacle is created on the way
of the exciton condensed phase pulse movement.

\begin{figure}[h!]
\includegraphics[width=\linewidth]{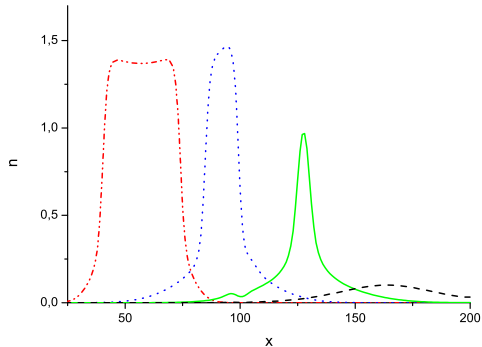}
  \centering
  \caption{Dependence of the exciton density on the $x$ coordinate at different moments of time in the presence of the barrier at $S=0.2$, $\Delta x=2.24$, , $\delta=0.09$, $b=4$, $P=0.64$, $\beta=10$, $\tau=1000$, $b_1=-2.23$, $d_1=0.2$.}
\label{fig:figure8}
\end{figure}

\begin{figure}[h!]
\includegraphics[width=\linewidth]{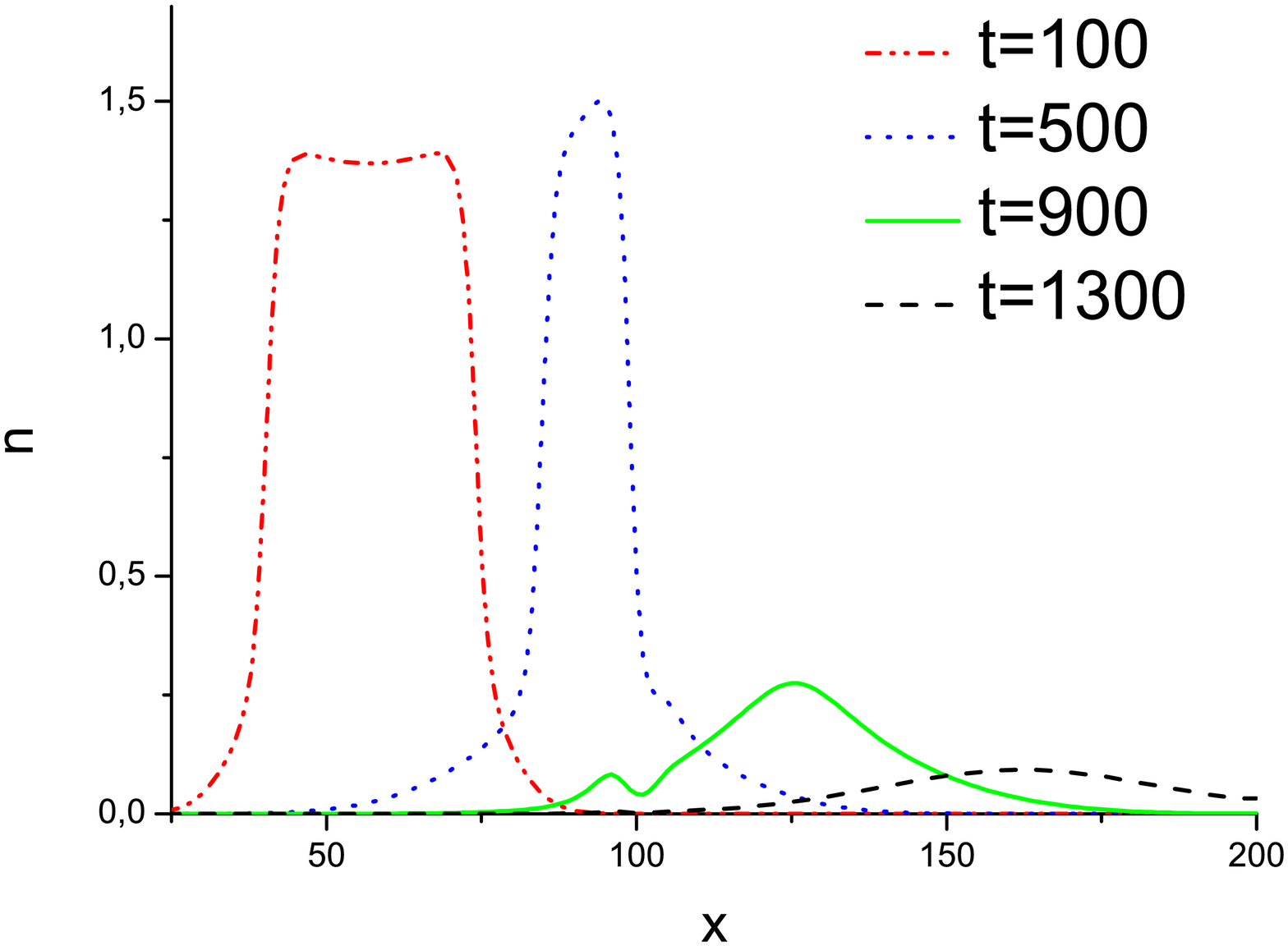}
  \centering
  \caption{Dependence of the exciton density on the $x$ coordinate at different moments of time in the presence of the barrier, $S=0.3$, $\Delta x=2.24$, , $\delta=0.09$,(other parameters are the same as in figure~\ref{fig:figure8}).}
\label{fig:figure9}
\end{figure}

\begin{figure*}[ht!]
  \vspace*{-3cm}
  \centering
  \begin{minipage}[b][7cm][b]{\linewidth}
    \begin{minipage}[t]{0.5\linewidth}
      \centering
      \includegraphics[width=\linewidth]{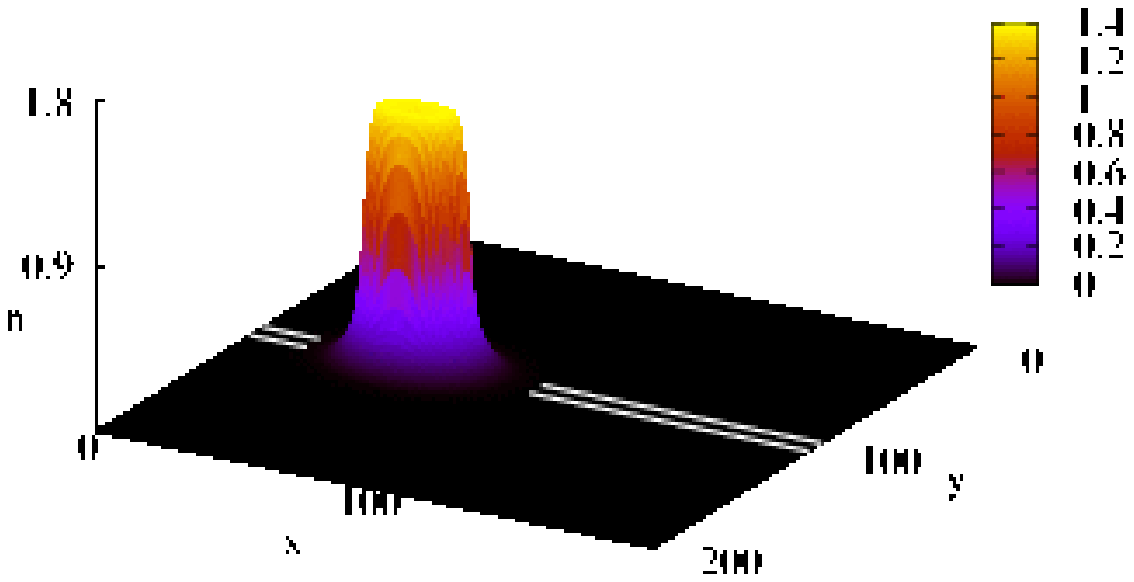}
      \par {\footnotesize {\bf a)} }
    \end{minipage}
    \hfill
    \begin{minipage}[t]{0.5\linewidth}
      \centering
      \includegraphics[width=\linewidth]{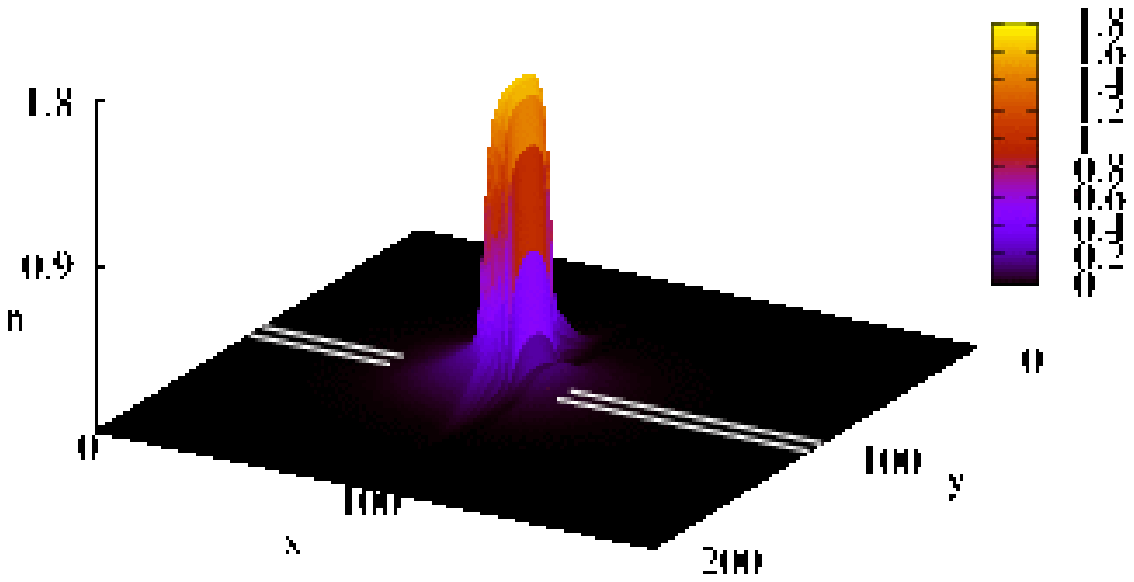}
      \par {\footnotesize {\bf b)} }
    \end{minipage}
  \end{minipage}
  \begin{minipage}[t][7cm][t]{\linewidth}
    \begin{minipage}[t]{0.5\linewidth}
      \centering
      \includegraphics[width=\linewidth]{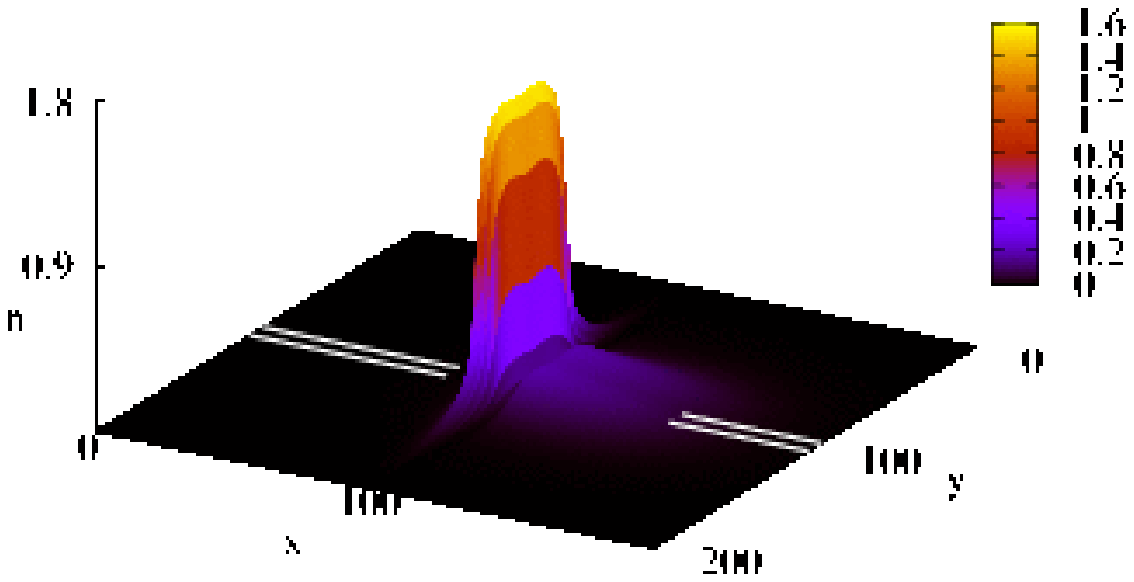}
      \par {\footnotesize {\bf c)} }
    \end{minipage}
    \hfill
    \begin{minipage}[t]{0.5\linewidth}
      \centering
      \includegraphics[width=\linewidth]{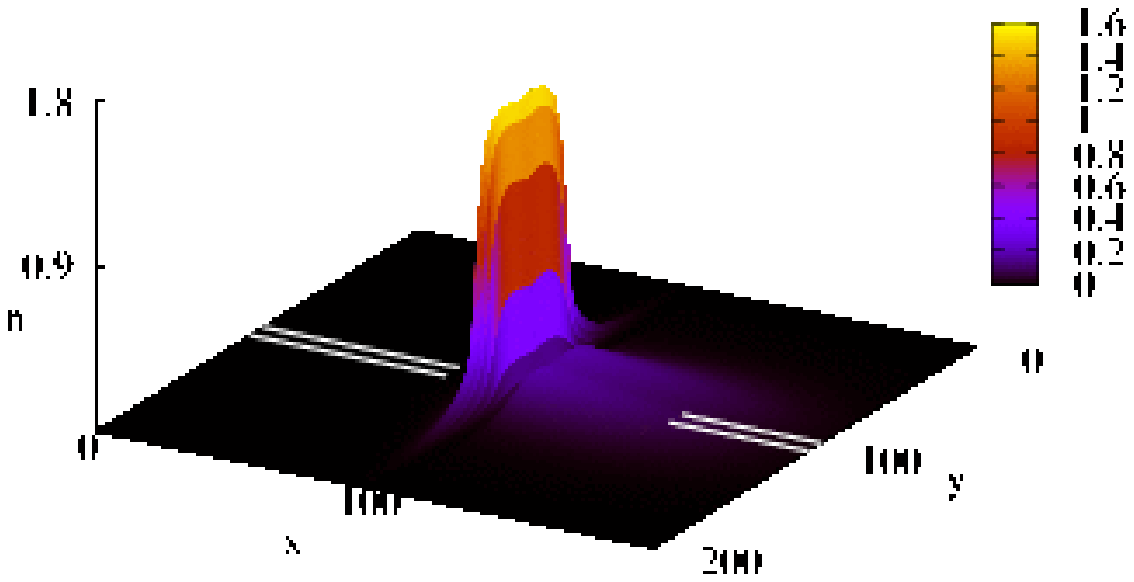}
      \par {\footnotesize {\bf d)} }
    \end{minipage}
  \end{minipage}
  \caption{Spatial distribution of the exciton density at $S=1$, $\Delta x=2.24$, $\delta=0.09$ (other parameters are the same as in figure~\ref{fig:figure8})  at different moments of time after switching off the pumping:a) $t=100$, b) $t=500$, c) $t=900$, d)$t=1300$. The white lines show the edges of the slot.}
  \label{fig:figure10}
\end{figure*}

\begin{figure}[h!]
\includegraphics[width=\linewidth]{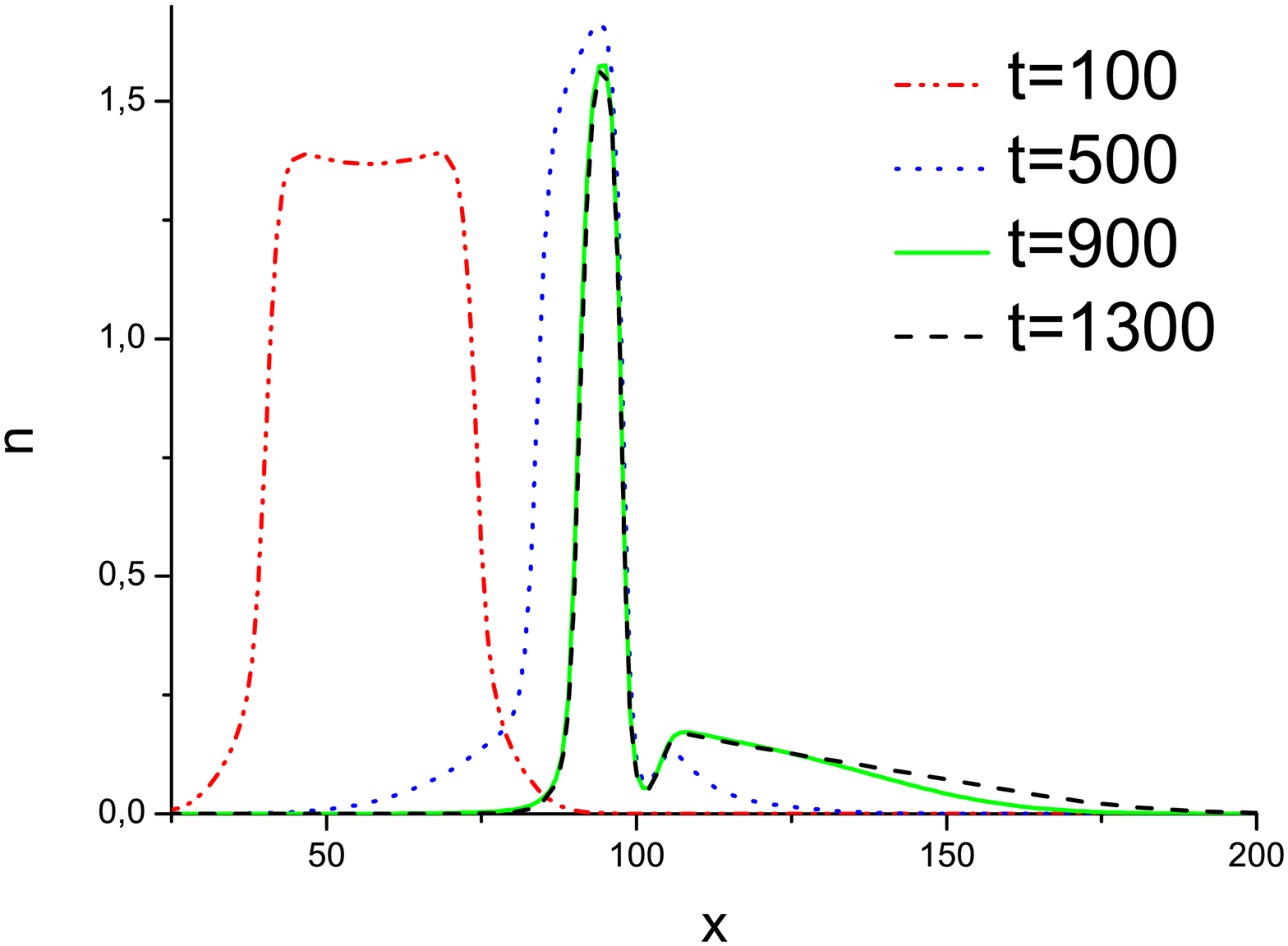}
  \centering
  \caption{Dependence of the exciton density on the $x$ coordinate at different moments of time in the presence of the barrier, $S=1$, $\Delta x=2.24$, $\delta=0.09$ (other parameters are the same as in figure~\ref{fig:figure8}).}
\label{fig:figure11}
\end{figure}

Let us do analysis of an influence of the height of the barrier on the penetration of the pulse through barrier. It is seen from comparison of the projections of the exciton density on the YZ plane for $S=0$ (figure~\ref{fig:figure5}) and $S=0.2$ (figure~\ref{fig:figure8}), that the magnitude of pulse amplitude drops after a barrier passing with increasing the height of the barrier. With increasing $S$ the exciton condensed phase pulse begins to destroy while passing through the barrier (figure~\ref{fig:figure9}) and the exciton condensed phase pulse exists beyond barrier only at $S\leq 0.3$. Dependence of the exciton density on spatial coordinates in the case of the presence of the barrier with parameters $S = 1$, $\Delta x=2.24$ and the position of the maximum $x_0 = 100$ is presented in two-dimensional case in figure~\ref{fig:figure10} and in the projection on YZ plane in figure~\ref{fig:figure11}. The position of the pulse does not change as a function of the time after the pulse reaches the barrier. So, the exciton condensed phase pulse stops in the area before the barrier and the exciton condensed phase pulse movement can be controlled by changing the barrier parameters. Simultaneously with stopping, the change of the shape of the pulse occurs (see figure~\ref{fig:figure10}).

Thus, the collective penetration of excitons through a barrier is similar to a quantum-mechanical particle tunnelling  through the barrier: the probability of the tunnelling decreases with the barrier growth. The situation with the dependence of passing processes on the width of the barrier (on $\Delta x$) is not unambiguous, since that considered by us potential for excitons consist of two parts: a linear part and the potential of the barrier. Depending on $\Delta x$ and $\delta$ the total height of the barrier may be decrease or increase as function of $\Delta x$ (see figure~\ref{fig:figure7}) with different effects on the pulse penetration.

So far we did calculation in dimensionless units. Let us do some quantitive estimations. For the quantitative description, we chose the following values of parameters: $\tau=100$~ns,
$T = 3.7$~K, $n_u=2 \cdot 10^{10}$ cm$^{-2}$, $D=9$ cm$^2$/s, $an_u=1.6$ meV.
 In this case the width of the slot is $2b = 5.36$ $\mu$m ($b=4$). Path traversed be the exciton condensed phase pulse during its lifetime is $l=67$ $\mu$m, and the height of the barrier is $S=an_u=1.6$ meV($S=1$).

\section{Exciton autosolitons}
So far we have studied the exciton density pulses localized in space. These pulses are created by laser pulses that act during some time interval. Exciton density pulses are damping over time. However, the existence of localized exciton pulses in space (autosolitons) is possible under stationary excitation radiation in some interval of the pumping~\cite{sugakov2013exciton}. Let us consider the possibility of formation of the exciton autosolitons in the potential created by the electrode with the slot. As it was shown in~\cite{sugakov2009ordered}, under the stationar irradiation in some interval of the pumping $G_{c1}<G<G_{c2}$ in the presence of the slot in the electrode periodical distribution of the exciton density in the quantum well arose. Homogeneous distribution of the exciton density exists outside this interval of pumping.

We showed that at steady-state pumping there is the spatial nonuniform solution
of the equation~\ref{equ:density_derivative_nodim} at
$G<G_{c1}$ in the form of an isolated peak~\cite{sugakov2013exciton}. It may be obtained solving the equation~(\ref{equ:density_derivative_nodim}) at the
pumping, which consist of a constant value $G_0$
and an
additional pulse $dG$
with the maxima in the some point
of the space and in the time moment 
\begin{equation}
dG=s\exp[-w((x-x_0)^2+(y-y_0)^2)]\exp[-p(t-t_0)^2], 
\label{equ:soliton}
\end{equation}
\noindent where $s, w, p$
are parameters. The equation~(\ref{equ:soliton}) describes
a pulse of the pumping, which acts during some time
interval with the maximum in the point $x_0, y_0$.

\begin{figure}[h!]
\includegraphics[width=\linewidth]{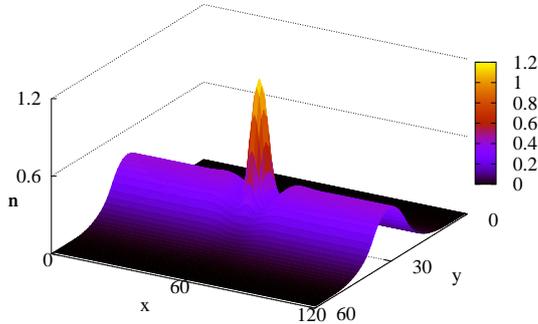}
  \centering
  \caption{Bright autosoliton. Spatial distribution of the exciton density at $G=0.005<G_{c1}$, $b=7$, $b_1=-2.23$, $x_0=60$, $y_0=30$.}
\label{fig:figure12}
\end{figure}

The solution of equation~(\ref{equ:density_derivative_nodim}) for the system of indirect excitons under the slot in the metal electrode obtained in the case of imposing
 the addition pulse (\ref{equ:soliton}) is presented in figure~\ref{fig:figure12}. The solution
exists at $t \rightarrow \infty$
, i.e. at the times, when the action
of the addition pulse is absent already. The shape of
the peak $n(x,y)$
does not depend on parameters
$s, w, p$ in some region of their values, but this solution exists in some region of pumping ($G=0.0048-0.0052$). In addition, the
solution in the form, presented in figure~\ref{fig:figure12}, arises also, if the
additional pulse is absent, but there is distribution
of the exciton density in the initial moment of time in the form:
\begin{equation}
n(x,y,0)=s\exp(-w((x-x_0)^2+(y-y_0)^2).
\end{equation}
Such localized spatial distribution of the exciton density is called "bright autosoliton". According to classification of~\cite{kerner1989autosolitons} it belongs to static soliton.

\begin{figure}[h!]
\includegraphics[width=\linewidth]{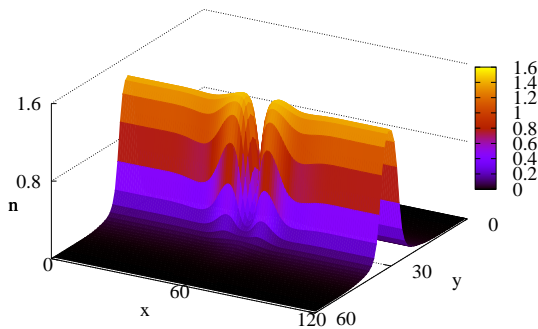}
  \centering
  \caption{Dark autosoliton. Spatial distribution of the exciton density at $G=0.008>G_{c2}$, $b=7$, $b_1=-2.23$, $x_0=60$, $y_0=30$.}
\label{fig:figure13}
\end{figure}

Localized solutions exist also in the some region at
the pumping greater the value, at which the periodical
structure arises ($G>G_{c2}$). The dependence of the
exciton density may be obtained from equation~(\ref{equ:density_derivative_nodim}) choosing
an additional pumping pulse in the form (\ref{equ:soliton}), but at
$s<0$. Such solution is presented in figure~\ref{fig:figure13}. These structures
can be called "dark autosolitons". They exist in some region of pumping($G=0.0078-0.0081$).
If there is an external field in a system,
which creates a linear potential (see~(\ref{equ:linear})), the autosoliton moves along the slot. So this autosolitons can be also used for information transmission in semiconductors. In this case the motion occurs under steady-state irradiation, but such pulse does not decay with time.

\section{Conclusions}

We consider the dynamics of the exciton condensed phase pulse in double quantum wells, created by the external irradiation. The model of the system
consists of a semiconductor structure sandwiched between two metal electrodes,
the top one having a slot that creates an additional potential for excitons.
The spinodal decomposition model, generalized to the case of particles with a
finite lifetime, was used to describe the exciton density distribution in condensed and gas phases.

To ensure movement of the exciton pulse, an additional potential that linearly
depends on the coordinates along the slot was included.  The control of the
exciton pulse is performed using both the barrier, which arises due to the
potential created by the metal strip, and the additional laser pulse. In
numerical calculations the following results were obtained:
\begin{enumerate}
\item Exciton condensed phase pulse arise if the laser pulse is greater than the
threshold value dependent on the parameters of the system,
\item In the case of the narrow slot, an exciton condensed phase pulse arise at
the centre of the slot, and in the case of the wide slot, an exciton condensed
phase pulse splits into two pulses, located near the boundaries of the slot.
\item In the presence of the potential that linearly changes along the slot, the
exciton pulse moves with the constant velocity. Maximum density is constant during the lifetime and the pulse does not spread if the exciton system is in the condensed phase.
\item Initial pulse can be amplified by imposing an additional laser pulse
thereby increasing the path traversed before attenuation,
\item Exciton pulse can be stopped or its shape can be changed by creating a potential barrier for excitons. With increasing the height of the barrier, passing of the excitons through the barrier decreases and the shape of the pulse changes. 
\item Existence of the bright and dark autosolitons in the system of excitons in double quantum well under the slot in the electrode was shown.
\end{enumerate}

The paper analyses peculiarities of collective effects, caused by phases formation, at exciton movement in inhomogeneous external fields. The experimental observation of the effects studied in this paper would allow
to determine the parameters of the exciton condensed phase and to investigate
the possibility of applying exciton systems in optoelectronics.

%\section*{References}

\bibliographystyle{unsrt}

\end{document}